\documentclass[preprint2]{aastex}
\usepackage{epsfig}
\def\micron{\mu{\rm m}}
\def\kms{ \, {\rm km s}^{-1}}

\include{macros}

\begin{document}

\title{The Clustering Dipole of the Local Universe from the 
Two Micron All Sky Survey}

\author{
Ariyeh H. Maller,
Daniel H. McIntosh,
Neal Katz and
Martin D. Weinberg
}
\affil{Astronomy Department, University of Massachusetts, Amherst, MA 01003}

\begin{abstract}
The unprecedented sky coverage and photometric uniformity of the
{\it Two Micron All Sky Survey} (2MASS) provides a rich resource for investigating 
the galaxies populating the local Universe.  A full	
characterization of the large-scale clustering distribution
is important for theoretical studies of structure formation.	
2MASS offers an all-sky view of the local galaxy population at 
$2.15\micron$, unbiased by young stellar light and minimally
affected by dust.  We use 2MASS to map the local 
distribution of galaxies, identifying the largest structures  
in the nearby universe. The inhomogeneity of these structures 
causes an acceleration on the Local Group of galaxies, which 
can be seen in the dipole of the Cosmic Microwave Background (CMB).
We find that the direction of the 2MASS clustering dipole is $11\arcdeg$
from the CMB dipole, confirming that the local galaxy distribution 
accelerates the Local Group.  From the magnitude of the dipole we find
a value of the linear bias parameter $b=1.37 \pm 0.3$ in the $K_s$-band.  
The 2MASS clustering dipole is $19\arcdeg$ from the latest 
measurement of the dipole using galaxies detected by the 
{\it Infrared Astronomical Satellite} (IRAS) suggesting that bias may be 
non-linear in some wavebands.
\end{abstract}


\section{Introduction}
Since the discovery of a dipole in the CMB \citep{lubin:83,fixs:83} it has 
been identified as the Doppler signature of our motion relative to the
CMB rest frame.  Strong evidence in favor of this interpretation has
been provided by the observation of a similar dipole in the surface brightness
of radio galaxies in the same direction \citep{bw:02}.  \citeauthor{bw:02}
make the distinction between velocity dipoles caused by our motion
and clustering dipoles which measure the distribution of galaxies.  Attempts
to measure the clustering dipole of local galaxies in the 1980s 
\citep{md:86,ywrr:86,lahav:87,vs:87} from only the fluxes and positions of 
galaxies led to results that were within $30\arcdeg$ from the CMB velocity
dipole.  The inclusion of redshift information led to a controversy
over out to which distance the clustering dipole converges 
\citep{sd:88,llb:89,rowan:90,pv:91,svz:91}. Through 
the 1990s progressively larger redshift surveys were used in an effort to 
determine where the clustering dipole converges and to improve the 
calculation of cosmological parameters 
\citep{stra:92,huds:93,schm:99,rowan:00}.
These improvements included determining the shot noise in the sample, 
optimizing the window function for smoothing and estimating the 
contribution of nonlinear effects to the velocity dipole \citep{stra:92}. 
These efforts have recently determined that the clustering 
dipole converges at $150 - 200 h^{-1}$ Mpc \citep{schm:99,rowan:00}.

2MASS \citep{skrut:97} is the first near-infrared ($JHK_s$ passbands) 
all-sky survey. 2MASS has an effective image resolution of 1 arcsecond
and $100\times$ greater sensitivity than the far-infrared all-sky survey, 
IRAS.  Most passbands tend to be sensitive to star formation rate,
while the $K_s$ passband is most sensitive to stellar mass \citep{bd:00},
which probably makes the $K_s$-band a better tracer of total mass.
Therefore, 2MASS offers
a unique opportunity for calculating a flux-weighted clustering dipole.

\begin{figure} [t] 
\centerline{\epsfig{file=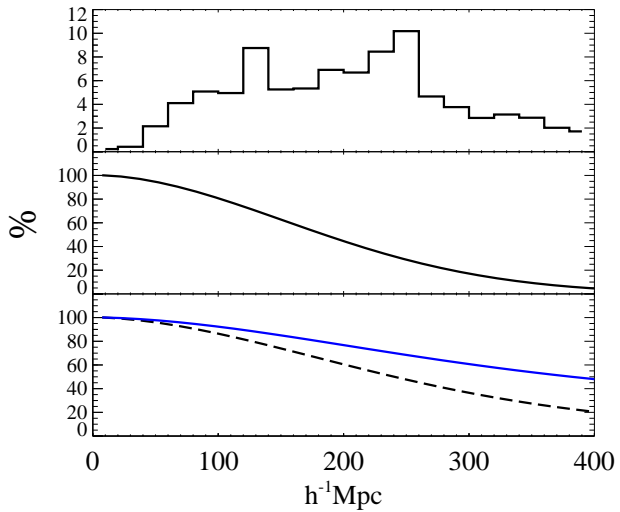,width=\linewidth}} 
\vspace{20pt}
\caption{Properties of the 2MASS galaxies as a function of distance $R$.
The top panel shows the percentage of 2MASS galaxies in $20 h^{-1}$ Mpc
bins as a function of distance.  The middle panel shows the percentage of 
luminosity that falls below our flux limit of $K_s = 13.57$ mag as a 
function of distance.  The bottom panel shows the percentage of the total 
integrated flux out to a given distance that is above the flux limit of 
$K_s = 13.57$ mag (solid line).  Also shown is the percentage of the total 
integrated luminosity (dashed line) as a function of distance, where both 
calculations have used the measured luminosity function \citep{bell:03}.
}\label{fig:select}
\end{figure}

The median depth of the survey is $z=0.073$, or $220 h^{-1}$ Mpc
\citep{bell:03}, a distance past where the dipole has been shown to converge. 
Figure \ref{fig:select} shows the distance distribution for a magnitude-limited
sample of 2MASS galaxies as 
determined from $6,104$ redshifts from the Sloan Digital Sky 
Survey early data release \citep{stou:02}.  
At a limiting magnitude of $K_s = 13.57$ mag, 2MASS includes 
all $L_*$ and brighter galaxies out to $200 h^{-1}$ Mpc. 

We can use the 2MASS luminosity function measured using the Sloan redshifts 
\citep{bell:03} to determine the completeness of the survey as a function 
of distance.  The result, shown in Figure \ref{fig:select}, is that the
survey contains $60\%$ of the total luminosity and $76\%$ of the total
flux inside $200 h^{-1}$ Mpc. Thus 2MASS is a fairly complete survey 
of the total flux in the local universe.  As we expect the clustering 
dipole to be dominated by bright objects, any missing flux is unlikely 
to contribute significantly to the dipole.  Later we show that this is 
true by demonstrating the convergence of the clustering dipole as a 
function of apparent magnitude.

\section{The 2MASS Catalog}
The 2MASS final release extended source catalog is the most complete 
resource of local galaxy observations at near-infrared wavelengths.  
The catalog contains over 1.6 million sources, and for most Galactic 
latitudes ($|b|>20\arcdeg$) has $>90\%$ completeness and $>98\%$ 
reliability down to $K_s\leq13.5$ mag \citep{jarr:00a}. 
For $5< |b| < 20\arcdeg$, in regions where the stellar density is less 
than $10^4$ per square degree with $K_s<14$ mag stars,
the catalog maintains its high completeness and is still 
$>80\%$ reliable \citep{jarr:00b}.  Hereafter, we adopt the \citet{kron:80} 
magnitudes, which attempt to recover the ``total'' galaxy flux using 
apertures related to the galaxy radius (limited in 2MASS to a 5\arcsec 
minimum).  
2MASS Kron magnitudes underestimate the true total flux systematically 
by $0.1$ mag for $K_s > 11$ \citep{bell:03}, which we correct for when 
converting magnitudes to flux. The extended source catalog is $97.5\%$ 
complete within the Sloan early data release for extinction-corrected Kron 
$K_s \leq 13.57$ mag \citep{bell:03}. While the 
reliability decreases at lower galactic latitudes we will adopt this limiting
flux and explore the effect of a brighter cutoff later on.  The photometry of 
the catalog is very uniform with errors of less then $0.1$ magnitudes 
\citep{niko:00}.  We have performed an analysis of the galaxy density with 
a number of possible contaminants and find that the cross-correlation 
is always less then $5\%$ \citep{mmkw:03b}.

We apply an extinction correction based on Galactic dust
maps \citep{sfd:98} to create a foreground corrected catalog of 
galaxies with $K_s < 13.57$ mag.  
From the overlap with the $6,104$ Sloan 
spectroscopically confirmed galaxies we find that bright ($K_s < 12$ mag)
galaxies have dust-corrected colors narrowly clustered around 
$0.8 < J-K < 1.2$.  Thus, we exclude objects brighter than
$K_s=12$ mag with $J-K < 0.6$ or $J-K > 1.4$, as there are no galaxies with 
these colors found in the confirmed sample ($\sim 2000$ objects).  
Since the photometric errors increase the spread in $J-K$ colors for 
faint objects we do not extend this color criteria below $K_s = 12$.

In the region $l > 230\arcdeg$ 
and $l < 130\arcdeg$ we mask out the region $|b| < 7\arcdeg$ and extend this 
to $|b| < 12\arcdeg$ in the region $l > 330\arcdeg$ and $l < 30\arcdeg$, where 
the 2MASS sources are clearly obscured.
We also remove $17$ galaxies that are members of the Local Group.
This leaves us with $748,080$ galaxies covering $90\%$ of the sky.  
Note that applying no dust correction reduces the number of galaxies 
to $704,035$.

\begin{figure*} 
\centerline{\epsfig{file=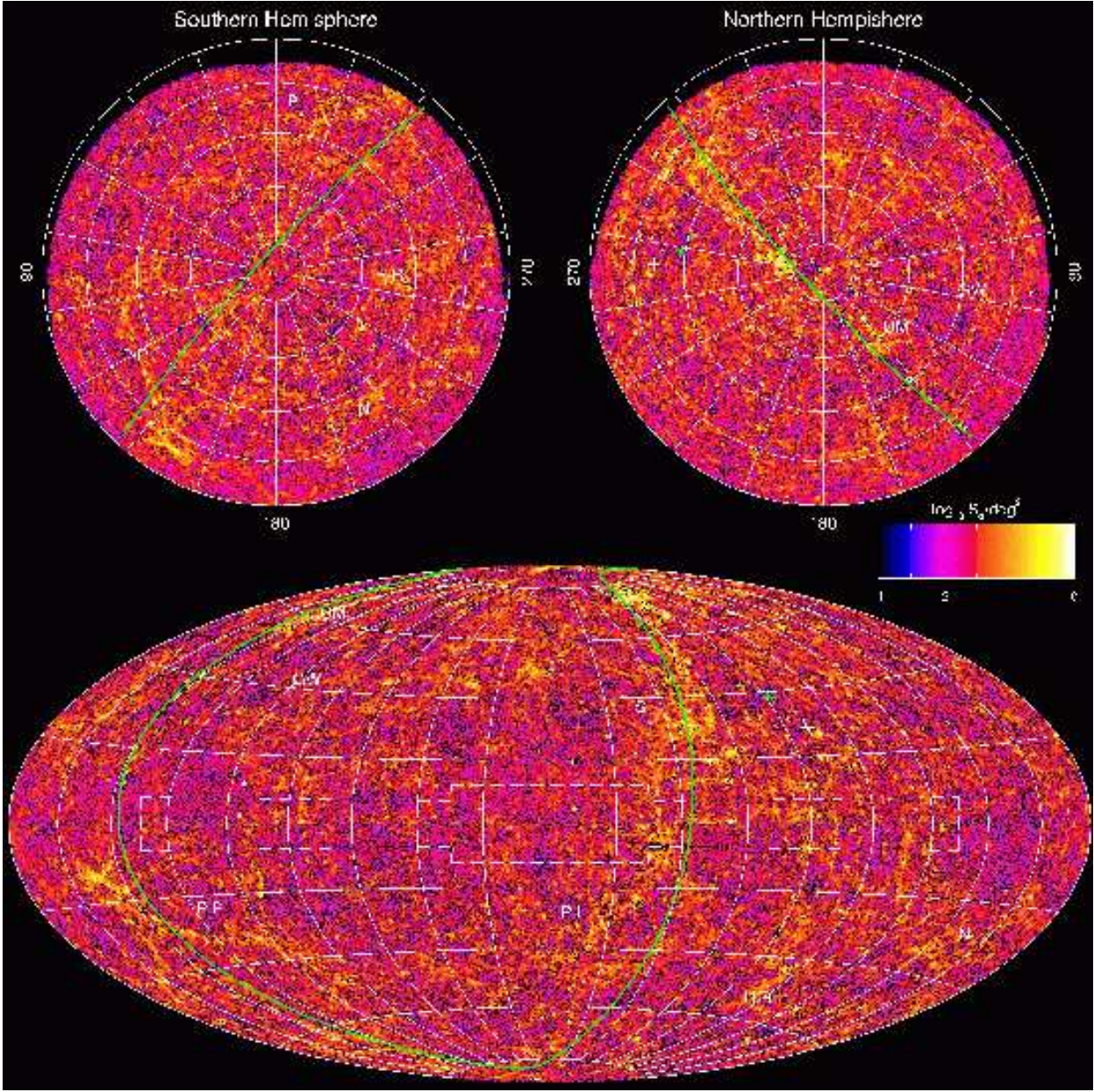,width=\linewidth}} 
\caption{The upper two images show the southern and northern Galactic 
hemispheres in Lambert's projection. The bottom image shows the whole sky
in an Aitoff projection. The color denotes the smoothed flux
density measured in units of the solar $K_s$-band 
flux per square degree. In the lower image the Galactic center 
has been masked out (dashed line) and adjacent regions have been cloned
to fill the masked region. We have labelled the most prominent
structures:  (S) ``Shapley'' complex, which includes 
Hydra-Centaurus in the foreground; (P-I) Pavo-Indus wall; (P-P) 
Perseus-Pisces chain; (H-R) Horologium-Reticulum supercluster;
(V) Virgo supercluster; (UM) Ursa Major cloud; (N) NGC 1600 group; and
(GW) the ``Great Wall''.  The green line delineates the super galactic 
plane along which many of these structures lie.  Also the CMB (white cross)
and 2MASS (green cross) dipoles are shown.  
}\label{fig:pic}
\end{figure*}

We display the most comprehensive representation of the local Universe
in Figure \ref{fig:pic}. We see a complex web of galaxies stretched out 
along filaments between dense nodes often corresponding to the location of 
giant superclusters. The most obvious structure is
the ``Great Attractor'' found within the region bounded by
$300<l<360\arcdeg$ and $-45<b<+45\arcdeg$.  This structure is
roughly divided into north and south Galactic hemisphere complexes -- the
overlapping Hydra-Centaurus and ``Shapley'' supercluster complex 
($0<b<+45\arcdeg$) and the Pavo-Indus wall ($-45<b<0\arcdeg$) 
\citep{dres:88,fisher:95}.
The Perseus-Pisces ``chain'' extends from the
Perseus supercluster ($l=150\arcdeg$,$b=-15\arcdeg$) to
($l=110\arcdeg$,$b=-35\arcdeg$)\citep{hg:86}, and appears to reach up 
to $b=+30\arcdeg$ along latitude $160\arcdeg$ \citep{fisher:95}.  
It has been demonstrated that there is no physical link 
between the Perseus-Pisces chain and the Pavo-Indus wall\citep{dinel:96}.
Additional features include the Horologium-Reticulum supercluster
($l=265\arcdeg$,$b=-55\arcdeg$) \citep{lucey:83}, Virgo supercluster
($l=285\arcdeg$,$b=+75\arcdeg$), the Ursa Major cloud
($l=145\arcdeg$,$b=+65\arcdeg$), and the NGC 1600 group
($l=210\arcdeg$,$b=-30\arcdeg$) \citep{fisher:95}.  Finally, 
the ``Great Wall'' at $cz\sim8000$~km/s \citep{gh:89} is the faint 
feature running from the Hercules supercluster 
($l=30\arcdeg$,$b=+45\arcdeg$) to ($l=180\arcdeg$,$b=+30\arcdeg$).

\section{Measuring the Clustering Dipole}
The inhomogeneity of these structures causes a gravitational acceleration 
on the Local Group of galaxies.  The CMB velocity dipole \citep{line:96}
results from the motion of the Sun relative to the CMB standard of rest.  
This motion can be broken down into two parts; the motion of the sun 
relative to the Local Group and the motion of the Local Group with respect 
to the CMB. Using the most recent values \citep{cv:99} we find that the 
Local Group velocity relative to the CMB is $622 \kms$ in the direction
$l_{cmb} = 272, b_{cmb} = 28$. 

Since both gravitational force and flux fall off as distance squared, the
net acceleration of the Local Group is proportional to the dipole
of the light distribution for a constant mass-to-light ratio $\Upsilon$.
Thus, 
\begin{equation}
\vec{g}=G \sum_i {{m_i \hat{r}_i}\over{r_i^2}}=G\Upsilon \sum_i S_i\hat{r}_i ,
\end{equation}
where $S_i$ is the flux received from each galaxy and $\hat{r}_i$ is
a directional unit vector. However, the light may be a biased tracer of
mass \citep{kais:84}; if there is linear biasing then the true acceleration 
would be a factor $1/b$ times the measured value.

Linear theory can be used to estimate the resultant velocity \citep{peeb:80},
which depends on the matter density of the universe $\Omega_{m}$. 
The expected velocity of the Local Group is therefore,
\begin{equation}
\label{eq:lin}
\vec{v} = \frac{2}{3} {{f(\Omega_m)}\over{b H_0\Omega_{m}}}\vec{g}.
\end{equation} 
where $f(\Omega_m) \simeq \Omega_m^{0.6}$ \citep{peeb:80} and $H_0$ is 
the Hubble constant. Of course, this is only true in linear theory. 
Nonlinear N-body simulations show that the 
velocity of a region like the local group is typically within $7\arcdeg$
of the computed linear acceleration and that the magnitude agrees to 
within $20\%$ \citep{dsy:91}. 
Measuring the light dipole therefore achieves two aims: (1) It verifies 
that the CMB dipole is truly caused by the Sun's motion and (2) it can
determine a combination of $\Upsilon_K$, $\Omega_{m}$ and $b$.
Since $\Omega_{m}$ and $\Upsilon_K$ can be measured by other means we 
can therefore determine the value of $b$.

To accurately calculate a clustering dipole we need to consider the effect of
the masked region.  
We address this in two ways: by cloning the sky above and below the masked
regions \citep{llb:89}, and by filling the masked region with randomly chosen 
galaxies such that it has the same surface density as the unmasked area.
We apply both methods to estimate the uncertainty contributed by 
the masked region.

\begin{figure} [t] 
\centerline{\epsfig{file=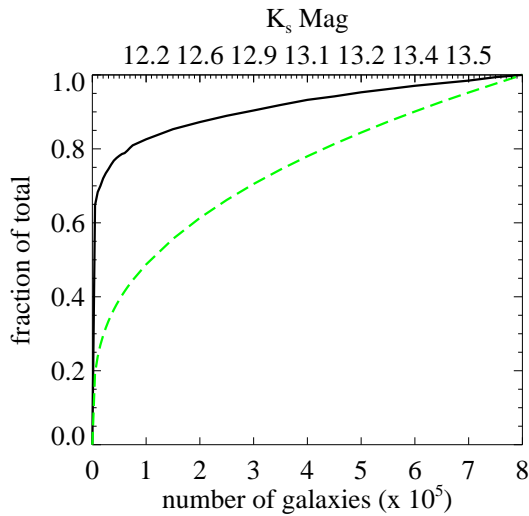,width=\linewidth}}
\caption{The growth of the clustering dipole is shown as a function
of the number of galaxies used to compute it ordered by their flux.  
The corresponding limiting $K_s$ magnitude is shown on the top axis.
The dipole quickly rises to $80\%$ of its final value.  The faintest 
$300,000$ galaxies only change its value by $< 5\%$.  The growth of 
the total flux (dashed line) for the same galaxies is shown for comparison.
The total flux rises more slowly and continues to rise even for the 
faintest magnitudes in our sample. 
}\label{fig:conv}
\end{figure}

Figure \ref{fig:conv} shows the convergence of the magnitude 
of the clustering dipole as a function of the flux limit of the survey.  
The dipole mostly arises from galaxies with $K_s < 12$ mag, the 
brightest $50,000$ galaxies.  The dipole converges at fainter 
magnitudes and the addition of $300,000$ galaxies with $K_s > 13.2$ changes 
the dipole by less then $5\%$.
For comparison, the dashed line shows the growth in the total flux as a
function of the number of galaxies used, which continues to grow for faint
magnitudes.  This is compelling evidence that the dipole has converged in
the 2MASS sample and that any flux we are missing in low luminosity galaxies
is a minor contribution.

To estimate the effect of shot noise on our calculation we perform bootstrap
resampling on the galaxy catalog.  We first bin the galaxies by their fluxes
and then resample each flux bin to ensure that the resampled catalog
has the correct flux distribution.  Performing the resampling $100$ times the
standard deviations of the clustering dipole direction is $\Delta l = 
0.5\arcdeg ,\Delta b = 0.3\arcdeg $ and in magnitude it is $0.5\%$.  We find
that the systematic uncertainties are much larger than the shot noise.

\begin{figure} [t] 
\centering 
\vspace{-30pt}
\epsfig{file=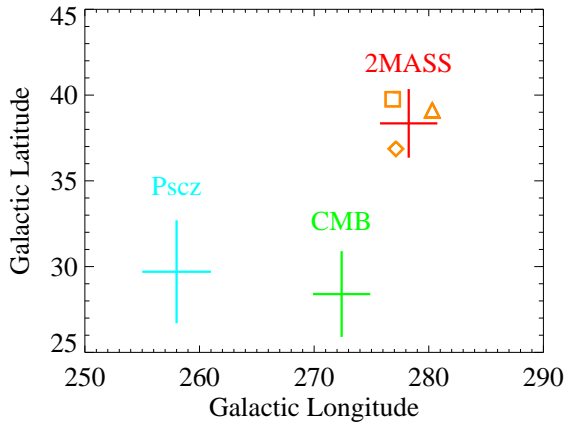,width=\linewidth} 
\vspace{-10pt}
\caption{The location of the 2MASS clustering dipole is shown relative to the 
CMB velocity dipole. Also shown is the location of the clustering dipole 
calculated from the IRAS Psc$z$ catalog \citep{rowan:00}.  
The position of the dipole with the masked area filled by cloning (triangle),
filled randomly (diamond), or including no dust correction (square)
are also shown.  The error bars reflect the 
systematic uncertainties caused by dust and the mask. 
}\label{fig:dip}
\end{figure}

The most important sources of systematic error in our calculation 
are our treatment of the mask and of dust corrections.  Cloning
the sky above and below the mask gives a dipole pointing towards
$l=280\arcdeg, b=39\arcdeg$. If instead of cloning the adjacent sky 
we fill the mask with randomly selected  galaxies the direction 
of the dipole changes to $l=277\arcdeg, b=37\arcdeg$.  
If we perform no dust correction on the galaxy catalog then the dipole 
moves to $l=277\arcdeg,b=40\arcdeg$ (Figure \ref{fig:dip}).  
We adopt the centroid of these three measurements 
$l_{dipole}=278 \pm 2.5 \arcdeg,b_{dipole}=38 \pm 2 \arcdeg$ as the best fit 
dipole with error bars reflecting the dust correction and mask filling 
uncertainties.  The dipole calculated only using galaxies brighter then 
$K_s = 13$ points to $l=278\arcdeg, b=40\arcdeg$ verifying that the missing
fainter galaxies are not introducing any significant error.
 
The clustering dipole is $11 \arcdeg$ from the CMB velocity dipole, 
closer then any previous determination.  This is somewhat larger then the 
average separation of $7\arcdeg$ found in the N-body simulations, but 
within $95\%$ confidence limits. Therefore we conclude that the 2MASS 
determined clustering dipole is consistent with the direction of the 
Local Group motion given the uncertainties caused by non-linear effects 
and systematic errors.  

Interestingly, our clustering dipole is $19\arcdeg$ from the dipole 
measured most recently using the IRAS Psc$z$ catalog \citep{rowan:00}.  
This is larger than the systematic errors in either calculation and 
therefore suggests differences either caused by the passband of the survey 
or in the methods used to calculate the clustering dipole. It is possible 
that biasing schemes more complicated then simple linear biasing may be 
needed to explain this disagreement.  The magnitude of the 2MASS clustering 
dipole is $1.25 \pm 0.06 \times 10^{-3} L_{\sun}$ kpc$^{-2}$, where again 
the errors are dominated by the systematic uncertainties in the dust 
correction and the treatment of the masked area.

\section{Estimating b}

The Wilkinson Microwave Anisotropy Probe \citep{benn:03} has recently 
determined the matter density of the universe to be $\Omega_m=0.27 \pm 0.04$
and $H_0 = 71 \pm 4 \kms {\rm Mpc}^{-1}$\citep{sper:03}.  From the 2MASS 
luminosity function we can determine the $K_s$ luminosity density of the 
universe to be $j_K=5.27\pm0.11 \times 10^8 h L_{\sun} Mpc^{-3}$ 
\citep{bell:03} and therefore the average mass-to-light ratio, 
$\Upsilon_K= 71 h^{-1} \pm 5 {{L_{\sun}}/{M_{\sun}}}$.
Using these values in eq. \ref{eq:lin} we find that the expected velocity
in linear theory is $856 b^{-1}\pm 80 \, \kms$ compared to the value measured 
from the CMB of $622 \, \kms$. Including the $20\%$ uncertainty between the 
linear prediction and the measured velocity in nonlinear N-body simulations 
we find $b=1.37 \pm .34$.  This can be compared with a measurement of 
bias from the 2df Galaxy Redshift Survey using redshift space distortions
which finds that $b =.94\pm.15$ in the $b_J$-band \citep{peac:01}. 

If bias is luminosity or color dependent then we would expect a higher 
value from 2MASS than from the 2df Galaxy Redshift Survey.
Since the measurement of bias as a function of waveband and scale is one
of the most important tools used to understand the formation and evolution
of galaxies, this measurement of the clustering dipole not only
confirms basic cosmological predictions, but will also lead to a better
understanding of how galaxies form.

\acknowledgements
We thank Roc Cutri, Tom Jarrett, Rae Stiening and Steve Schneider for
help with the 2MASS catalog.  We thank Micheal Struass for a useful critic
of an earlier draft of this work.  This publication makes use of data products 
from the Two Micron All Sky Survey, which is a joint project of the 
University of Massachusetts and the Infrared Processing and Analysis 
Center/California Institute of Technology, funded by the National 
Aeronautics and Space Administration and the National Science Foundation

\bibliographystyle{apj}
\bibliography{dipole,2mass,me,cosmo}

\end{document}